\documentclass[prd,superscriptaddress,preprintnumbers,twocolumn,showpacs]{revtex4-1}
\usepackage{times}
\usepackage{amsmath}
\usepackage{graphicx}
\usepackage{amssymb}
\usepackage{latexsym}
\usepackage{bm}
\usepackage[normalem]{ulem}  
\usepackage[dvips]{color} 

\renewcommand\sout{\bgroup \color{red} \ULdepth=-.5ex \ULset}
\renewcommand{\sout}[1]{}

\newcommand{\ssB}{{\scriptscriptstyle B}}
\newcommand{\muB}{\mu_{\ssB}}
\newcommand{\dmu}{\delta\mu}

\newcommand{\MeV}{\mathrm{MeV}}

\newcommand{\TCP}{T_\mathrm{CP}}

\newcommand{\nn}{\nonumber}
\begin{document}
\preprint{KUNS-2447 / YITP-13-26}
\title{QCD phase diagram at finite baryon and isospin chemical potentials\\
in Polyakov loop extended quark meson model with vector interaction
}
\author{H. Ueda}
\affiliation{Department of Physics, Faculty of Science, Kyoto University,
Kyoto 606-8502, Japan}
\affiliation{Yukawa Institute for Theoretical Physics, Kyoto University,
Kyoto 606-8502, Japan}
\author{T. Z. Nakano}
\affiliation{Department of Physics, Faculty of Science, Kyoto University,
Kyoto 606-8502, Japan}
\affiliation{Yukawa Institute for Theoretical Physics, Kyoto University,
Kyoto 606-8502, Japan}
\author{A. Ohnishi}
\affiliation{Yukawa Institute for Theoretical Physics, Kyoto University,
Kyoto 606-8502, Japan}
\author{M. Ruggieri}
\affiliation{Department of Physics and Astronomy, University of Catania,
Via S. Sofia 64, I-95125 Catania, Italy}
\author{K. Sumiyoshi}
\affiliation{Numazu College of Technology,
Ooka 3600, Numazu, Shizuoka 410-8501, Japan}

\begin{abstract}
We investigate the QCD phase diagram of isospin asymmetric matter using
the Polyakov loop extended quark meson (PQM) model with vector interaction.
The critical point temperature is found to decrease
in isospin asymmetric matter
and disappear at large isospin chemical potential.
We also discuss the QCD phase transition in the neutron star core.
From comparison of the QCD phase diagram in PQM
and corresponding baryon and isospin chemical potentials
of neutron star matter in relativistic mean field models,
we show that the order of the chiral phase transition in the neutron star core
could be crossover because of large isospin chemical potential.  
\end{abstract}

\pacs{12.38.Lg, 21.65.Qr}

\maketitle

\section{Introduction}
The QCD phase transition would be realized not only in heavy-ion
collisions but also in compact astrophysical objects and phenomena such as heavy
neutron stars~\cite{Haensel:2007yy}, supernovae~\cite{SN} and black hole formations~\cite{Sumiyoshi:2006id,Sumiyoshi:2007pp,Sumiyoshi:2008kw,Ohnishi:2011jv}.
At zero baryon chemical potential ($\muB$), the QCD phase transition
at finite temperature ($T$) is
accessible by using the lattice Monte Carlo simulation,
e.g.~\cite{Karsch:2001cy,Greensite:2003bk}. At large $\muB$ , $\muB /T \gtrsim 1$,
the situation is much less clear since the lattice simulation is plagued by
the well-known sign problem~\cite{LQCD-finite-mu}.
We can investigate this region by using chiral effective models such as
the Nambu--Jona-Lasinio (NJL) model~\cite{NJL}
and the quark meson (QM) model~\cite{QM},
and the chiral effective model with the Polyakov loop effects such as
the Polyakov loop extended Nambu--Jona-Lasinio (PNJL)
model~\cite{Meisinger:1995ih,Fukushima:2003fw,Ratti:2005jh}
and the Polyakov loop extended quark meson (PQM)
model~\cite{Schaefer:2007pw,Skokov:2010sf}.
The QCD phase diagram, especially the QCD critical point (CP) location,
strongly depends on models and model parameters~\cite{Stephanov:2007fk}.
Therefore, further experimental and theoretical developments are necessary
to determine the structure of the QCD phase diagram.

For laboratory experiments, the search for CP in heavy-ion collisions is
ongoing at RHIC~\cite{BES} and is planned in the coming FAIR facility.
Since the phase transition is second order at CP, the coherence length $\xi$
is divergent and large fluctuations of the order parameter are expected
in a volume of the size $\xi ^3$.
Various signatures of CP have been proposed
theoretically~\cite{Sig-CEP}.
It is not an easy task to observe the divergence signature of
$\xi$ in heavy-ion collisions, since the system size and the evolution
time are limited.
Moreover, it is difficult to create cold dense matter, and CP may not be
reachable in the laboratory if CP is located in the high density region,
$\muB>500~\MeV$.

By comparison, very dense matter is formed in compact astrophysical
phenomena. For example, high density and low
temperature matter is formed in the neutron star core, and high
temperature and high density matter is produced during a gravitational
collapse of a massive star and binary stars~\cite{Binary_NS}.
From the observation of these phenomena, we may get information on the QCD
phase diagram in the high density region~\cite{Haensel:2007yy,SN,Sumiyoshi:2006id,Sumiyoshi:2007pp,Sumiyoshi:2008kw,Ohnishi:2011jv}.
In compact astrophysical phenomena, charge neutrality leads to
suppressed proton fraction compared with that
of neutrons, and the isospin chemical potential $\dmu \equiv
(\mu_n -\mu_p)/2 = (\mu_d - \mu_u)/2$ is finite and positive.
In particular, $\dmu$ appears as another
independent thermodynamical variable in supernovae and BH formations,
since trapped neutrinos modify the neutrinoless charge neutrality condition $(\dmu = \mu_e /2)$.
Therefore it is necessary to consider $\dmu$ dependence of the QCD
phase diagram in order to discuss the QCD phase transition in compact
star phenomena.

The phase structure in the three thermodynamic
variables $(T,\mu ,\dmu)$ is still an open problem.
In our previous work~\cite{Ohnishi:2011jv},
we have discussed the possibility of the CP sweep during BH formation
processes where $\delta \mu$ is finite;
quark matter core and hadronic envelope may merge to one phase,
when the temperature exceeds the CP temperature ($T_\mathrm{CP})$.
The location of CP strongly depends on $\dmu$;
for large $\dmu$, $T_\mathrm{CP}$ becomes lower
and it becomes more probable
for the heated matter to go through CP. 
%
There are several recent works which discuss the QCD phase diagram
in charge neutral dense matter~\cite{PNJL-neutral}
and in the three-dimensional space,
$(T,\mu,\dmu)$~\cite{PNJL-dmu,FRG} or $(T,\mu,\mu_L)$~\cite{PNJL-lepmu},
where $\mu_L$ is the lepton-number chemical potential.
The phase diagram structures in these works have some differences.
In Ref.~\cite{PNJL-neutral},
the isospin chemical potential is found to be small $\delta\mu<m_\pi/2$
in charge neutral quark matter, and pions are not found to condense.
In Ref.~\cite{PNJL-dmu},
three-dimensional $(T,\mu,\delta\mu)$ phase diagram is investigated
in the mean-field approximation,
and the $s$-wave pion condensed phase is found to appear
in the finite $\mu$ and $\delta\mu$ region.
$\TCP$ decreases with increasing $\delta\mu$,
until the CP hits the pion condensation phase boundary.
In Ref.~\cite{FRG},
fluctuation effects are taken into account in the quark meson model
by using the functional renormalization group (FRG) flow equation.
$\TCP$ is also found to decrease with increasing $\dmu$.
The $s$-wave pion condensed phase is found
in the high $\dmu$ and low $\mu$ region,
but it is suppressed at large $\mu$.
As a result, the pion condensed phase is separated from
the chiral first order phase transition surface 
in the $(T,\mu,\delta\mu)$ space.
In Ref.~\cite{PNJL-lepmu},
$T_\mathrm{CP}$ is found to be insensitive to the lepton-number
chemical potential.

In this article, we investigate the isospin chemical potential
dependence of the QCD phase diagram in more detail, and discuss the order
of the chiral phase transition in the neutron star core, where $T=0$ and
$\muB ,\dmu > 0$.
For this purpose, we first compute the QCD phase diagram using
the two-flavor PQM with vector interaction
and examine the $\dmu$
dependence of the QCD phase diagram.
According to the $s$-wave $\pi N$ repulsion argument~\cite{spion}
and functional renormalization group
results~\cite{FRG}, we assume that pions do not condensate.
We then discuss the order of the chiral phase transition in neutron star
core, through the comparison of the QCD phase diagram with the $\beta$
equilibrium line in neutron star matter calculated by using hadronic
equation of states (EOSs). 

The article is organized as follows: In Sec.I\hspace{-.1em}I,
we briefly describe PQM with vector interaction.
The results are discussed in Sec.I\hspace{-.1em}I\hspace{-.1em}I, where
we show the $\dmu$ dependence of the QCD phase diagram and compare
the QCD phase diagram in PQM with neutron star matter chemical potentials.
Sec.I\hspace{-.1em}V is devoted to summary and discussion.   
\section{Polyakov loop extended quark meson model }
%
\subsection{PQM Lagrangian and parameters}
In this Section, we describe the PQM model 
augmented with the vector
interaction.
PQM is an effective model which has the chiral symmetry and confinement
property of QCD~\cite{Schaefer:2007pw,Skokov:2010sf}.
The Lagrangian density of the two-flavor PQM is given by~\cite{Schaefer:2007pw,Skokov:2010sf}
\begin{align}
{\cal L} &= \bar{q}\left[
i\gamma^\mu D_\mu
- g(\sigma + i\gamma_5\bm\tau\cdot\bm\pi)
- g_\omega\gamma^\mu\omega_\mu - g_\rho\gamma^\mu\bm\tau\cdot\bm{R}_\mu
\right]q 
\nonumber\\
&+ \frac{1}{2}(\partial_\mu\sigma)^2 +
\frac{1}{2}(\partial_\mu\bm\pi)^2 -U(\sigma,\bm\pi)
\nonumber\\
&- \frac{1}{4} \omega_{\mu\nu}\omega^{\mu\nu}
- \frac{1}{4} \bm R_{\mu\nu} \cdot \bm R^{\mu\nu}
+ \frac{1}{2} m_v^2 (\omega_\mu\omega^\mu
	+ \bm R_\mu\cdot \bm R^\mu)
\nn\\
&-{\cal U}(P,\bar P,T) ~,
\label{eq:PQM} 
\end{align}
where $q$ denotes a quark field with Dirac, color and flavor indices,
$\bm\tau$ is the Pauli matrix in the flavor space and $\omega^{\mu
\nu}$ and $\bm R^{\mu \nu}$ are the field tensors of $\omega $ and
$\rho$ mesons.
The mesonic potential $U$ and the Polyakov loop potential
$\mathcal{U}$ are given as,
\begin{align}
 U(\sigma,\bm\pi) &= \lambda ( \sigma^2 + \bm \pi^2 - v^2)^2/4 -
 h\sigma ~,
\label{eq:Meson}\\
 \mathcal{U}[P,\bar P,T] &= T^4\biggl\{-\frac{a(T)}{2}
 \bar P P + b(T)\ln H(P,\bar{P})
 \biggr\} \;,\label{eq:Poly}\\
 H(P,\bar{P})&= 1-6\bar PP + 4(\bar P^3 + P^3) -3(\bar PP)^2 ~.
\end{align}
$\sigma$ and $\bm \pi$ are
the isoscalar-scalar and isovector-pseudoscalar meson fields.
The covariant derivative $D_\mu=\partial_\mu - iA_\mu$ in Eq.~\eqref{eq:PQM} is the Dirac operator with a temporal static and
homogeneous background gluon field $A_\mu = \delta_{\mu 0}A_0$.
Without the
explicit symmetry breaking term, the last term in Eq.~\eqref{eq:Meson}, the
Lagrangian in Eq.~\eqref{eq:PQM} has $SU(2)_L \times SU(2)_R$ symmetry.

$\mathcal{U}(P,\bar P, T)$ is an
effective potential of gluon field,
where $P$ and $\bar P$ are the Polyakov loop and its
conjugate,
\begin{align}
 P = \frac{1}{N_c}\text{Tr}L~,\hspace*{1cm} \bar P = \frac{1}{N_c}\text{Tr}L^\dagger~.
\label{eq:PolDef}
\end{align}
$L$ is defined in the Euclidean space as,
\begin{align}
 L = {\cal P}\exp\left(i\int_0^\beta d\tau A_4\right),
\end{align}
where ${\cal P}$ stands for the path ordering.
The logarithmic term $\ln H({P,\bar P)}$ in Eq.~\eqref{eq:Poly}
comes from the Haar measure of the
group integral in strong-coupling lattice QCD~\cite{Fukushima:2003fw}. Coefficients, $a(T)$ and $b(T)$, are given
as functions of $T$, and parameterized as
$a(T)= a_0+a_1 (T_0/T)+a_2 (T_0/T)^2$,
and $b(T)=b_3 (T_0/T)^3$~\cite{Ratti:2005jh}.
\subsection{Effective Potential}

We now give the effective potential in dense asymmetric matter in PQM.
In asymmetric matter, $u$ and $d$ quark populations are unbalanced,
and we need to
introduce two independent chemical potential for $u$ and $d$ quarks
\begin{align}
 \mu_u = \mu - \dmu~, ~~~ \mu_d = \mu + \dmu~,
\label{eq:chemical}
\end{align} 
where $\mu =\muB /3$ is the quark chemical potential.
The isospin chemical potential $\dmu$ is
an independent thermodynamical variable in supernovae or black hole formation
processes,
while the neutrino-less $\beta$-equilibrium condition, $\dmu=\mu_e/2$,
applies to cold neutron star matter.

We assume that the $\sigma$ meson
and the temporal components of $\omega$ and $\rho^0$ mesons
take finite expectation values,
while others do not.
These expected values are assumed to be constant.
In this approximation, the quark single-quasiparticle
energy is given by
\begin{align}
 E_{fp}^* &= E_p + g_\omega \omega + g_\rho \tau^3 R~,
\label{eq:Efp}
\end{align}
 with
\begin{align}
 E_p = \sqrt{\bm p ^2 + M^2},~~ M= g\sigma~.
\end{align}
$\omega$ and $R$ in Eq.~\eqref{eq:Efp} denote the expectation values
of $\omega$ and $\rho^0$ mesons ($\omega=\langle \omega_0
\rangle$,$R=\langle R_0^3 \rangle$), respectively, where the
subscript 0 denotes the temporal component and the superscript for $R$
shows isospin.
The effect of vector interaction is to shift the quark chemical
potential~\cite{Vector-Int}. For later convenience, we define effective chemical potentials
for $u$ and $d$ quarks,
\begin{align}
 \tilde \mu_u = \mu -\dmu - g_\omega \omega - g_\rho R\ ,~~
 \tilde \mu_d = \mu +\dmu - g_\omega \omega + g_\rho R\ .
\end{align}
Integrating over the quark fields results in the following effective potential,
\begin{align}
 \Omega_{PQM}&= \mathcal{U}(P, \bar P, T) + U(\sigma, \bm\pi = 0) +
 \Omega_0 + \Omega_T~,\\
 \Omega_0 &= -2N_f N_c\int \! \frac{d \bm p}{(2\pi)^3}E_p\theta
 \left(\Lambda ^2 - \bm p^2\right)~,\label{eq:Vac}\\
 \Omega_T &= -\frac{1}{2}\left(m_\omega^2 \omega^2 +
 m_\rho^2R^2 \right) 
\nn\\
&-2T\sum _f \int \! \frac{d\bm
 p}{(2\pi)^3}\log{\left(F_-^f F_+^f\right)}~,\\
 F_-^f &= 1+3 P e^{-\beta {\cal E}_-^f } + 3 \bar P e^{-2\beta{\cal
 E}_-^f} +e^{-3\beta{\cal E}_-^f}~,\label{eq:FMdef}\\
 F_+^f &= 1+3\bar P e^{-\beta {\cal E}_+^f } + 3Pe^{-2\beta{\cal
 E}_+^f} +e^{-3\beta{\cal E}_+^f}~, \label{eq:FPdef}\\
 {\cal E}_{\pm}^f &= E_p \pm \tilde \mu_f ~,
\end{align}
where $\Omega_T$ is the thermal contribution
and $\Omega_0$ is the fermion vacuum energy, regularized by the
ultraviolet cutoff $\Lambda$. This term is necessary to reproduce the
second-order chiral phase transition at zero baryon chemical potential
$\muB$ in the chiral limit~\cite{Skokov:2010sf}.
Each term on the right hand side of Eq.~\eqref{eq:FMdef} corresponds to
the thermal contribution of zero, one, two, and three quark states.
Similarly, Eq.~\eqref{eq:FPdef} is the thermal contribution of antiquarks.
While PQM is
renormalizable and we can use dimensional renormalization~\cite{Skokov:2010sf},
it is sufficient to cut large momenta by a hard cutoff
for our purposes.

The equations of motion are obtained from the stationary conditions
in equilibrium,
\begin{align}
 \frac{\partial \Omega}{\partial \sigma} = 
 \frac{\partial \Omega}{\partial P} =
 \frac{\partial \Omega}{\partial \bar P} =
 \frac{\partial \Omega}{\partial \omega} =
 \frac{\partial \Omega}{\partial R}
 = 0~
 .
\end{align}
We obtain $(T,\muB,\dmu)$ dependence of the mean fields,
$\sigma, P, \bar{P}, \omega$ and $R$,
by solving these equations.  

Here we do not consider the pion condensation because the s-wave pion
condensation will not occur in neutron stars
when we take account of the s-wave $\pi N$
repulsion~\cite{spion}. Functional renormalization group analysis
also shows the shrinkage of the pion condensed region at finite $\mu$ 
than naively expected ($\dmu>m_\pi$/2)~\cite{FRG}.
%
\subsection{Model parametrization}

The parameter in the scalar-pseudoscalar part, $g,\lambda,\nu, h$ are fixed to
reproduce some properties of quarks and mesons in vacuum for a given
value of the hard momentum cut off $\Lambda = 600~$MeV in this work.
The quark-scalar meson coupling $g$ is determined by the constituent quark
mass in the vacuum $m_q = g\sigma = 335~\rm{MeV}$.
The mesonic potential parameters $\lambda$ and $ v$ 
are given by the chiral condensate in the vacuum 
$\sigma = f_\pi = 92.4~$MeV, 
and the $\sigma$ meson mass $m_{\sigma}^2 =\partial^2 \Omega/\partial
\sigma^2= (700~\rm{MeV})^2$.
The explicit symmetry breaking
parameter $h$ is  given by the pion mass $h = m_{\pi}^2f_{\pi}$.

In this study, we assume the quark-vector
couplings are the same ($g_{\omega} = g_{\rho} = g_v$) for simplicity.
We regard  $g_v$ as a free parameter, and we compare the results with  several values of
$r = g_v/g$.
We also assume the common vector meson masses$(m_\omega= m_\rho = m_v =770~\rm{MeV})$ 

The parameters in the Polyakov loop potential are fitted to the pure gauge lattice data~\cite{Boyd:1996bx}.
The standard choice of the parameters reads~\cite{Ratti:2005jh}
$a_0 = 3.51, a_1 = -2.47, a_2 = 15.2$ and $b_3 = -1.75$.
The parameter $T_0$ in Eq.~\eqref{eq:Poly} sets the deconfinement
scale in the pure gauge theory, i.e. $T_0 = 270$ MeV.
Chemical potential dependence of these parameters 
is not considered in this work~\cite{Schaefer:2007pw,Fukushima:2010is}.

\section{Results}
%
In this section, we discuss the $\dmu$ and the vector
coupling dependence of the QCD phase diagram.
The chiral phase transition is found to be weakened
at finite $\dmu$ or with finite vector coupling $r$.
In order to demonstrate this point,
we first discuss the order parameters as functions of $\muB$
at several values of $\dmu$ and $r$.

\begin{figure*}[bth]
 \begin{center}
  \includegraphics[width=7cm]{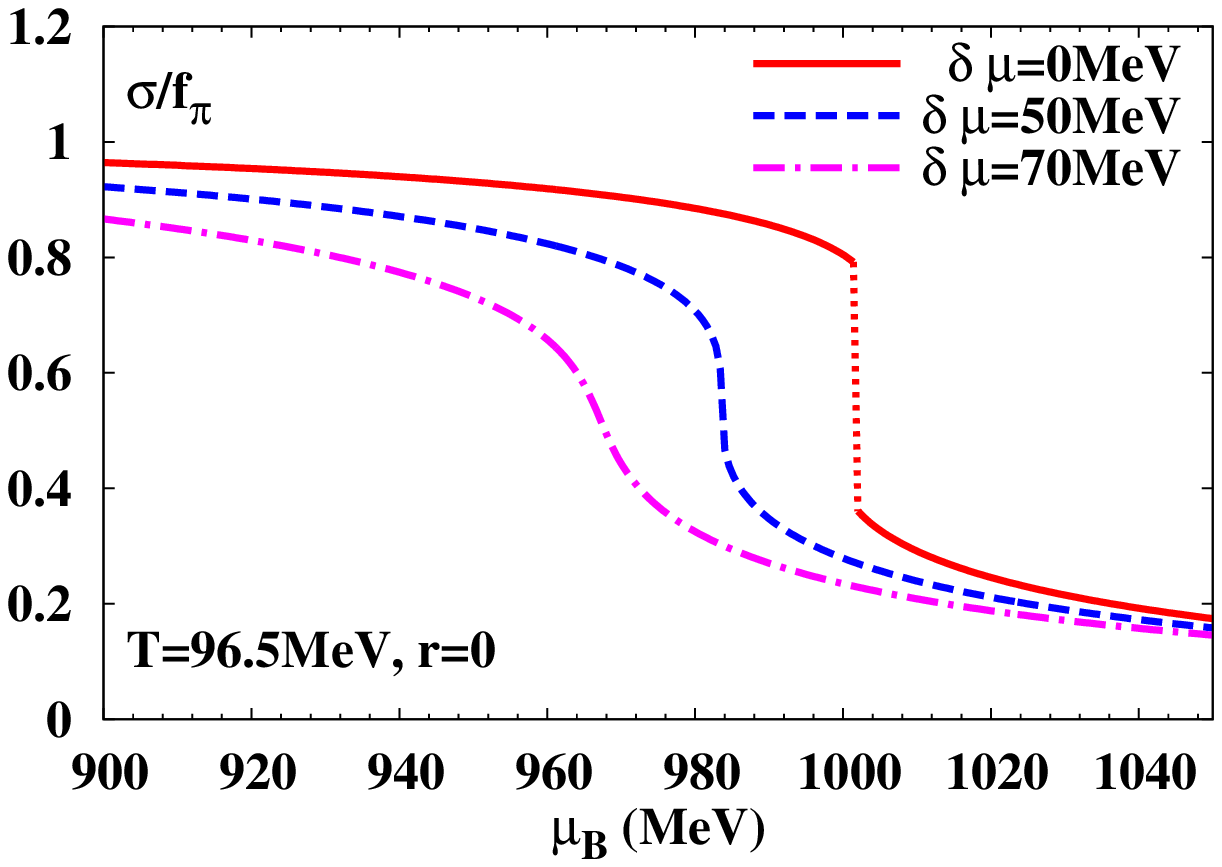}
  \includegraphics[width=7cm]{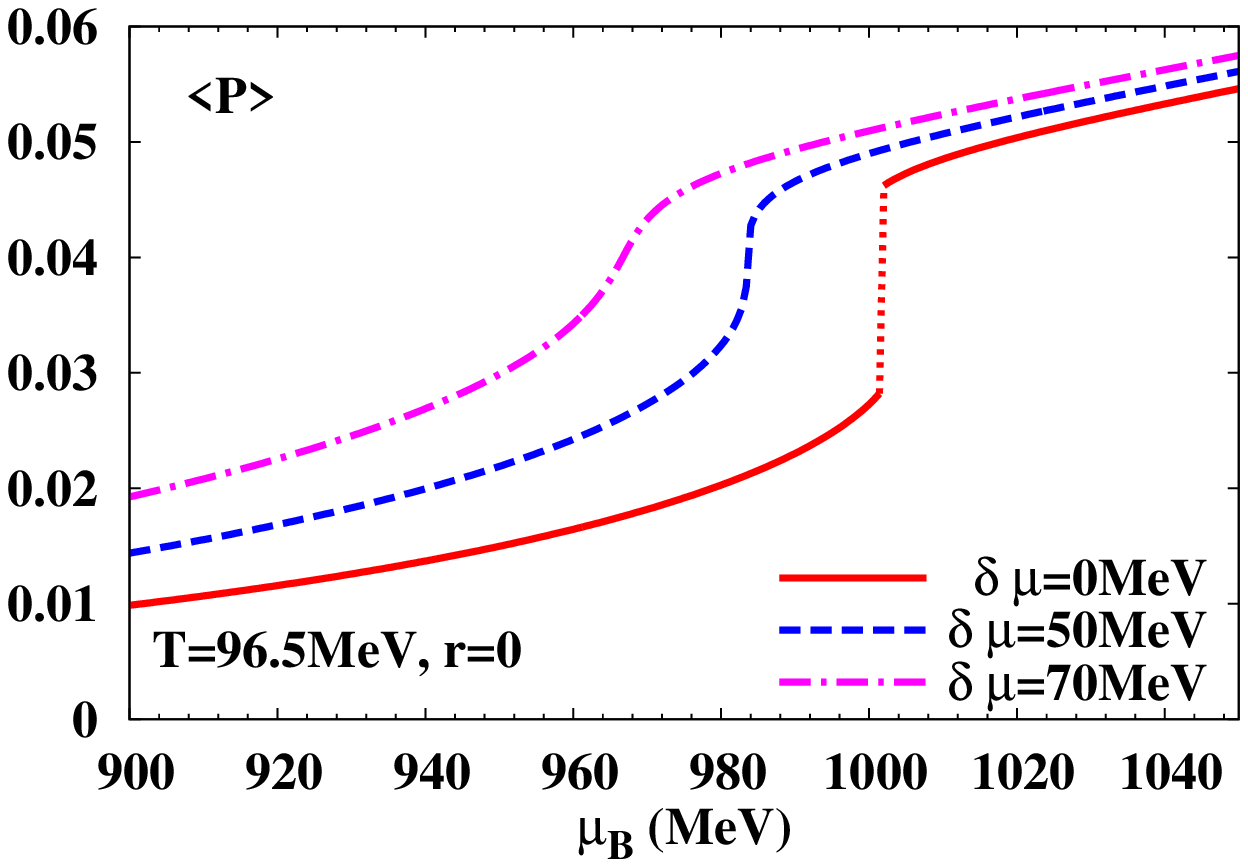}
\end{center}
 \caption{The order parameters $\sigma$(left panel), $P$(right panel)
 as functions of 
 baryon chemical potential $\muB$
 at 
 $T = 96.5~\MeV =T_\mathrm{CP}(\dmu=50~\mathrm{MeV}, r=0)$
 and three different isospin chemical potentials
 $\dmu = 0$ (solid line), 50 (dash line), 70 MeV (dash-dot line).
 The vector-scalar coupling ratio is chosen to be $r=0$.
}
\label{fig:order_dmu}
\end{figure*}

The phase structure is 
obtained from
the behavior of the order parameters $\sigma, P$ and $\bar P$.
Figure~\ref{fig:order_dmu} shows $\muB$ dependence of the order
parameters, $\sigma$ (left) and $P$ (right), for several isospin
chemical potentials at
$T = 96.5~\MeV =T_\mathrm{CP}(\dmu=50~\mathrm{MeV}, r=0)$
(CP temperature at $\dmu = 50$~MeV
and the vector-scalar coupling ratio $r=0$).
For small $\dmu$, the chiral phase transition is
first-order, while for $\dmu \gtrsim$ 50MeV, the chiral phase
transition becomes crossover. The change of the nature
of the phase transition with the increase of $\delta\mu$ is not a peculiarity
of the PQM model; in fact, several chiral 
models share this property, as discussed in~\cite{Ohnishi:2011jv} (see also the Appendix for
a discussion within the NJL model).

\begin{figure*}[bth]
 \begin{center}
  \includegraphics[width=7cm]{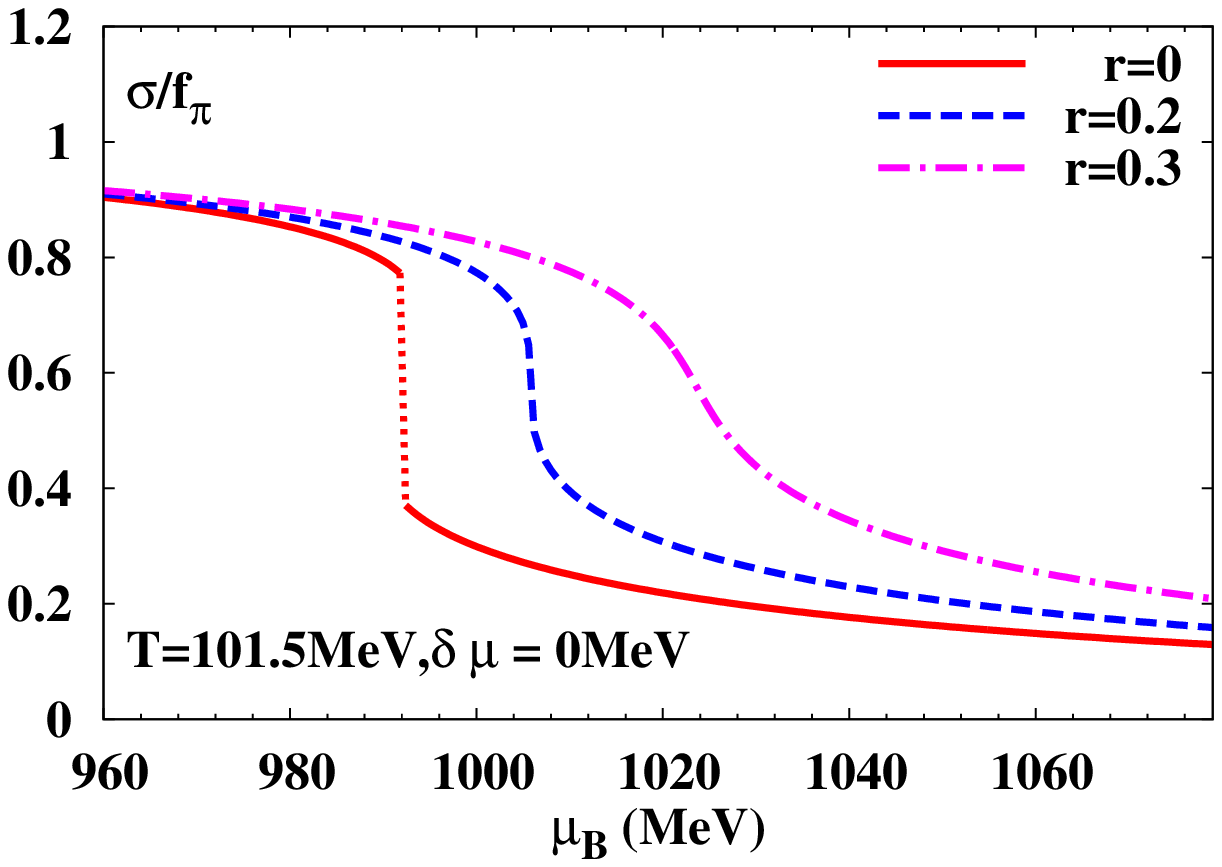}
  \includegraphics[width=7cm]{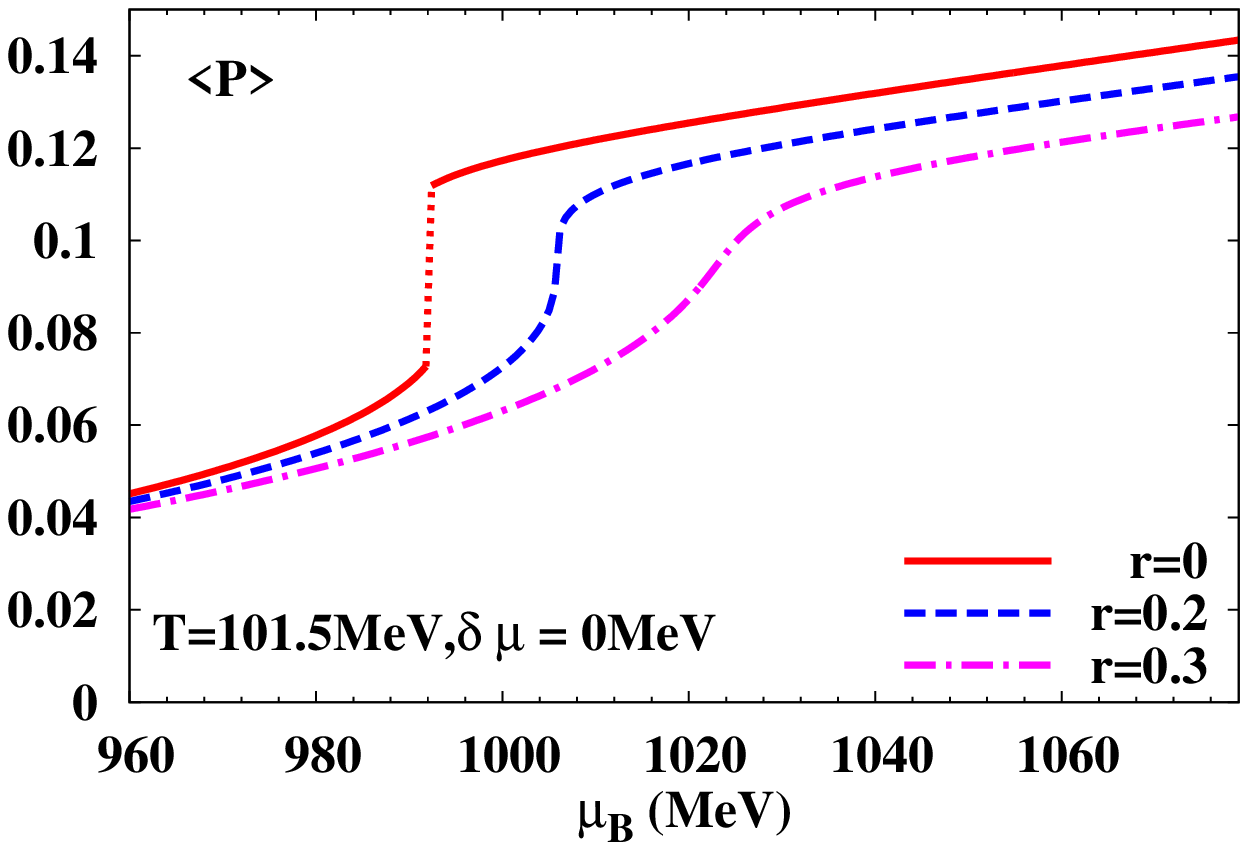}
\end{center}
 \caption{The order parameters  $\sigma$(left panel), $P$(right panel)
 as functions of baryon chemical potential $\muB$
 at $T = 101.5~\MeV =T_\mathrm{CP}(\dmu=0~\mathrm{MeV}, r=0.2)$
 and $\dmu=0~\mathrm{MeV}$
 for several values of the vector-scalar coupling ratio $r=0,0.2,0.3$.
 }
\label{fig:order_v}
\end{figure*}

In Fig.~\ref{fig:order_v}, we show $\sigma$ (left) and $P$ (right)
as functions of the baryon chemical potential at $T = 101.5~\MeV =
T_{\rm CP}(\delta \mu =0, r=0.2)$
and $\dmu=0~\MeV$
for several values of the vector-scalar coupling ratio $r$.
For the strong vector interaction,
the transition chemical potential is shifted to higher values, and
the chiral phase transition is smoothed.
The transition becomes crossover
for $r \gtrsim 0.2$ at this $T$.

\begin{figure}[bth]
 \begin{center}
  \includegraphics[width=7.5cm]{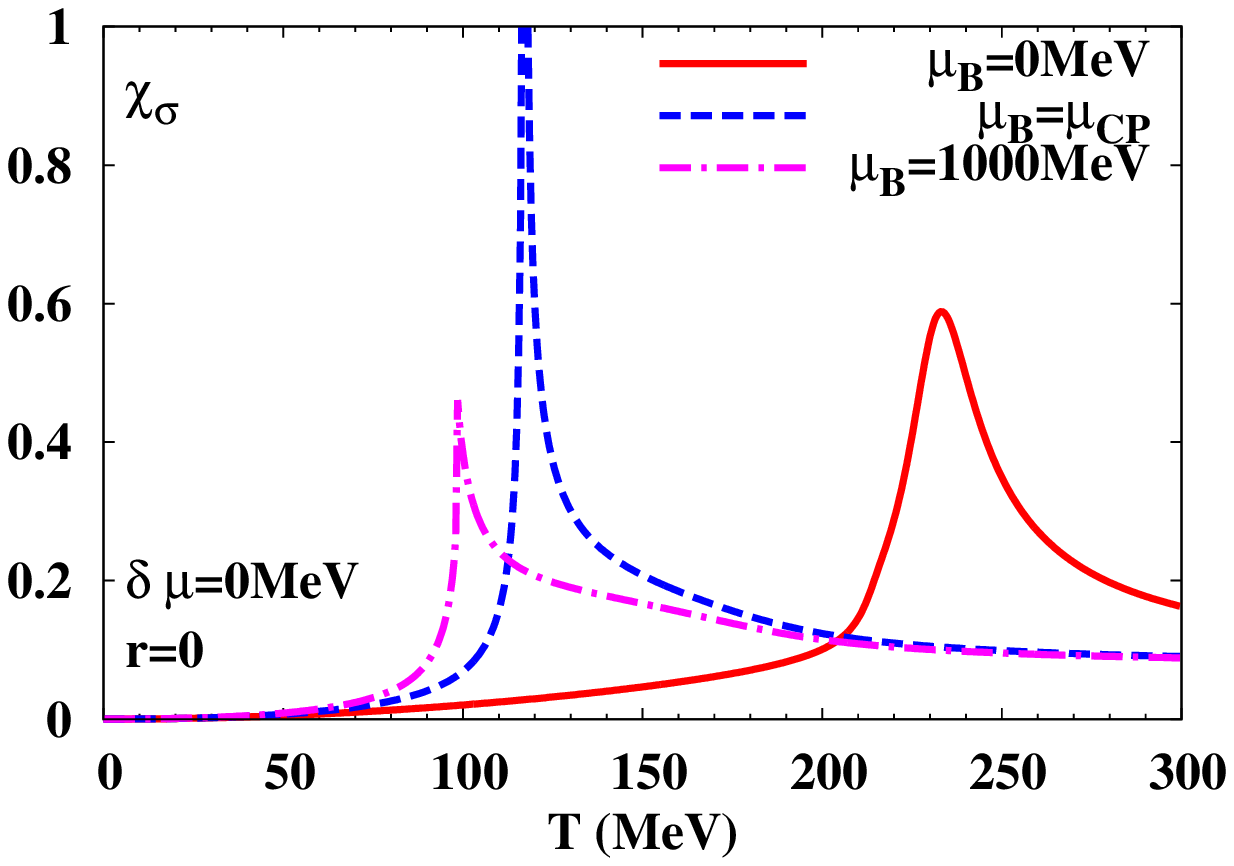}
 \end{center}
 \caption{The chiral susceptibility $\chi_{\sigma}$ as a function of
  temperature for different isospin chemical potential $\muB =
  0$(solid line), $\mu_\mathrm{CP}$(dash line), 1000~MeV(dash-dot line) at
 $\dmu =0~$MeV.  $\chi_{\sigma}$ is
  divergent at the critical end point.}
 \label{fig:chi_sig}
\end{figure}

We next discuss the $\dmu$ and vector coupling dependence of
the chiral and deconfinement phase boundaries.
Since the chiral phase transition at small $\muB$ is actually a smooth crossover
for finite quark masses, we have to establish a criterion to identify the phase boundary
of the chiral transition.
Here we define the chiral
critical temperature $T_c$ or the baryon chemical potential $\mu_{\ssB,c}$
of the chiral phase transition
by the peak of the chiral susceptibility $\chi_{\sigma}$
as a function of $T$ or $\muB$
for fixed $\dmu$ and $\muB$ or $T$, respectively.
Since $\chi_{\sigma}$ is divergent at the critical point, 
we can unambiguously determine the critical point temperature
$T_\mathrm{CP}$ and baryon chemical potential $\mu_\mathrm{CP}$
by the diverging peak of $\chi_\sigma$
in the $T-\muB$ plane. 
The chiral susceptibility 
is defined as the second derivative of
the effective potential by the explicit chiral breaking coefficient $h$,
\begin{align}
 \chi_{\sigma} = - T^3\frac{\partial^2 (\Omega/T)}{\partial h^2}~.
\label{eq:chi_sig}
\end{align}
Since $h \propto m_\pi^2$ is proportional to the bare quark mass,
the above definition 
gives a susceptibility which is proportional
to the usual definition around the critical point,
$\chi_{\sigma} = - \partial^2 \Omega / \partial M^2 / T^2$,
where $M$ is the bare quark mass.
We normalize Eqs.~\eqref{eq:chi_sig}
by multiplying some powers of $T$
to consider dimensionless susceptibility.
Figure \ref{fig:chi_sig} shows the chiral susceptibility as a function of
temperature for several baryon chemical potentials
at $\dmu = 0~\MeV$.
For each $\muB$, we find a peak in $\chi_{\sigma}$,
where the chiral phase transition occurs.
At CP, 
$(T,\muB)=(T_\mathrm{CP}, \mu_\mathrm{CP})$,
this quantity is divergent which signals a second order
phase transition. The critical points are found to be
$(T_\mathrm{CP},\mu_\mathrm{CP}) = (117,975)~\MeV$
at $\dmu=0$ without vector coupling $r=0$.

As for the case of the chiral transition, the deconfinement transition is a crossover 
for finite quark masses,
and we need to specify a criterion to identify the deconfinement phase boundary.
Several prescriptions to define the critical temperature
for deconfinement have been used in the literature: 
the temperature at which the Polyakov loop susceptibility, $\chi_P$, is maximum;
the temperature at which $dP/dT$ is maximum~\cite{Kahara:2008yg,Miura:2011kq};
finally, the half-value prescription, 
in which one identifies the deconfinement temperature with the average 
of the temperatures at which at $P=1/2$ and 
$\bar{P}=1/2$~\cite{Kahara:2010wh} (the two differ at finite $\mu$).
The Polyakov loop susceptibility $\chi_P$ 
may have a double peak structure
in some cases~\cite{Fukushima:2003fw}:
one peak is related to the chiral phase transition
and the other is related to the transition caused
by the Polyakov loop mean field potential.
A similar double peak behavior is found
in $dP/dT$~\cite{Kahara:2008yg,Miura:2011kq}.
Thus it is not easy to unambiguously define the deconfinement temperature
from the Polyakov loop susceptibility or the temperature derivative.
Since the Polyakov loop is small ($P, \bar{P} \simeq 0$)
in confined phase and large ($P, \bar{P} \simeq 1$) in deconfined phase,
the half-value prescription is the simplest one to adopt.
Therefore, we adopt the half-value prescriptions to define the deconfinement temperature.

\begin{figure}[bth]
 \begin{center}
  \includegraphics[width=7.5cm]{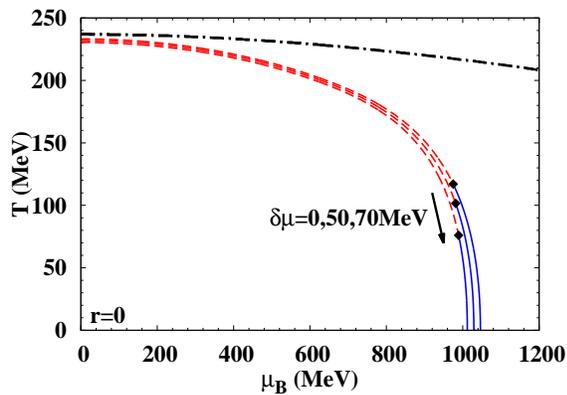}
 \end{center}
 \caption{The QCD phase diagrams for several isospin chemical potential.
 The red dash and the blue solid lines show the crossover and first order
 chiral phase transition boundaries respectively at
 $\dmu = 0,50, 70~\MeV$. The square dots show the CP.  The black dash-dotted lines show the
 confinement-deconfinement phase boundaries at $\dmu = 0, 70$MeV.
 }
 \label{fig:PD_dmu}
\end{figure}

We show the QCD phase boundaries for several $\dmu$ values
in Fig.~\ref{fig:PD_dmu}.
The hadron phase shrinks a little
and the critical point temperature $T_\mathrm{CP}$ decreases
with increasing $\dmu$,
while the confinement-deconfinement phase boundary
only weakly depends on $\dmu$.
The reduction of the transition chemical potential may be understood
as the density effects.
For a simple estimate,
let us consider the low $T$ transition in the chiral limit
without the vector coupling,
where the sum of $u$ and $d$ quark number densities in the chiral restored phase
is proportional to
$(\mu+\dmu)^3+(\mu-\dmu)^3=2\mu^3(1+3\dmu^2/\mu^2)$ as in the free
massless case .
If the QCD phase transition at finite $\dmu$ occurs
at the same density in the Wigner phase as that for $\dmu=0$,
the transition quark chemical potential is calculated to be
$\mu\simeq\mu_c-\dmu^2/\mu_c$,
where $\mu_c$ represents the transition chemical potential at $\dmu=0$.
This estimate gives the transition chemical potential shifts of
7.2 and 14 MeV for $\dmu=50$ and 70 MeV, respectively,
which is comparable to the PQM results, 7.0 and 13 MeV.
Another possible explanation is the decrease of the effective number of flavors.
At finite $\dmu$, one of the $u$ or $d$ quarks is favored,
and the phase diagram is expected to be closer to that at $N_f=1$,
where the phase transition is weaker.
%
%
%

We note that the deconfinement transition temperature $(T_d)$
is a little higher than the chiral transition temperature $T_c$.
The present behavior is consistent with the lattice Monte-Carlo
simulation results, which suggest
$T_d > T_c$~\cite{LQCDT}.
It should be noted that this order
depends on the choice of $T_0$ and is different from some of
the effective model results~\cite{Schaefer:2009ui}.
While the order of $T_d$ and $T_c$ at $\mu=0$ is an interesting problem
on the relation of deconfinement and chiral transitions,
it is irrelevant to our conclusion
and we choose $T_0=270~\MeV$ in the later discussion.

It should be noted that the deconfinement phase boundary is almost insensitive
to the baryon chemical potential, leading to a splitting of the chiral
and deconfinement transition boundaries.
This behavior is similar to the strong coupling lattice QCD results
including finite coupling and Polyakov loop effects~\cite{Miura:2011kq},
but it is different from the results obtained from the functional
renormalization group method starting from the PQM initial condition
at large cutoff~\cite{Pawlowski}.

\begin{figure}[bth]
 \begin{center}
  \includegraphics[width=7.5cm]{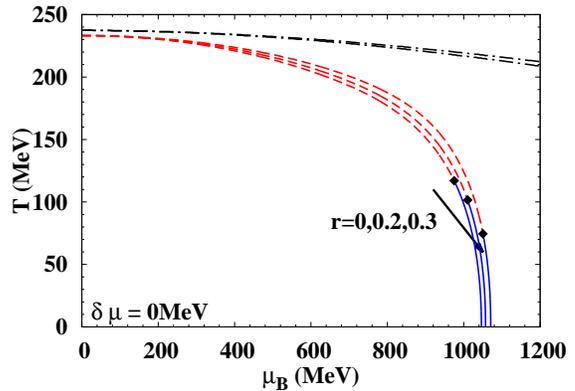}
 \end{center}
 \caption{The QCD phase diagrams for several quark-vector meson couplings.
 The red dash and blue solid lines show the crossover and first order
 chiral phase transition boundaries respectively at $ r =
 0,0.2,0.3$. The black dash-dotted lines show the
 confinement-deconfinement phase boundaries at $r = 0$ and $r=0.2$.
 }
 \label{fig:PD_r}
\end{figure}

Figure~\ref{fig:PD_r} shows the QCD phase diagrams
of symmetric matter ($\dmu=0$)
for several quark-vector meson couplings.
With increasing vector coupling,
the chiral phase boundary moves to the higher $\muB$ direction,
and the CP moves to the higher $\muB $ and lower $T$
direction.
The behavior of $\mu_\mathrm{CP}$ is
understood from the effective $\muB$ shift.
We can ignore
the $\rho^0$ meson
effects in symmetric matter, and
the effective chemical potential is given as
$\tilde{\mu} = \mu -rg\omega$.
Therefore, a strong vector interaction makes $\tilde{\mu}$ small for a given $\muB$~\cite{Fukushima:2008wg},
and the phase boundaries and the CP moves to high $\muB$
for finite vector coupling, $r\not=0$. 
By comparison, the vector coupling dependence
of the confinement-deconfinement phase boundary is small.

\begin{figure}[htbp]
 \begin{center}
  \includegraphics[width=7.5cm]{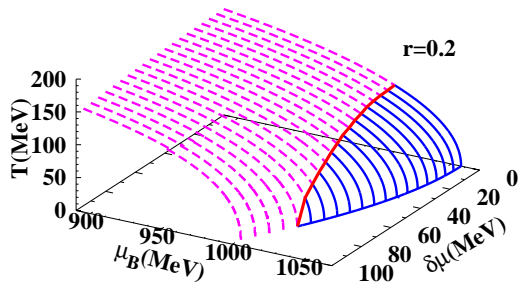}
 \end{center}
  \caption{The 3-dimensional ($T,\muB,\dmu$) QCD phase diagram in
 PQM with $r=0.2$. The dash lines and the blue solid lines show the
 crossover and first order chiral phase boundary respectively. The red
 solid line shows CP for different $\dmu$.}
 \label{fig:PD3D}
\end{figure}

\begin{figure}[htbp]
 \begin{center}
  \includegraphics[width=7.5cm]{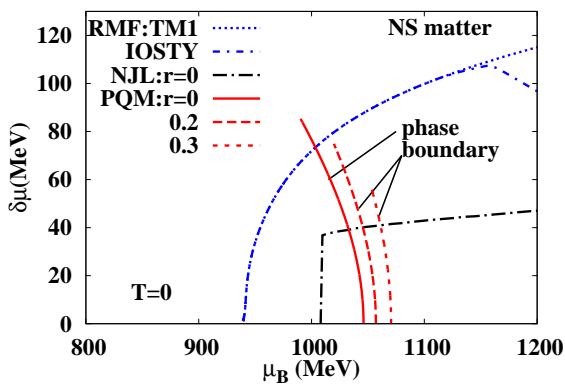}
 \end{center}
 \caption{This figure shows the first order phase boundaries at $T=0$ in PQM
 model with $r=0$(red solid line), 0.2(red dash line) and 0.3(red dot line)
 and $(\muB, \dmu)$ for neutron star matter calculated by
 NJL~\cite{Abuki:2009ba} with $r=0$(black dash-dot line),
 TM1~\cite{Sugahara:1993wz}(blue dot line) and IOSTY~\cite{Ishizuka:2008gr}(blue dash-dot line)~\cite{spion}.
 }
 \label{fig:PDT0}
\end{figure}

We show the
QCD phase diagram in $(T, \muB, \dmu)$ space in Fig.~\ref{fig:PD3D}.
As already mentioned,
$\dmu$ reduces $T_\mathrm{CP}$
and the transition baryon chemical potential at $T=0$.
Then the first order boundary narrows with increasing $\dmu$,
and eventually
the CP disappears at a certain value of $\dmu$.
This happens also for the NJL model, as we discuss in more detail in the Appendix.
This behavior is important when we consider the chiral phase transition in 
dense and isospin asymmetric matter, which is realized in
the core of neutron stars, where $\dmu$ becomes large.
For example, the reduced CP temperature may affect
the dynamical black hole (BH) formation processes.
The highest temperature during the BH formation
is calculated to be $T \sim 70~\mathrm{MeV}$,
and compressed matter may experience either
the first order, crossover, or CP sweep depending on the CP location
in asymmetric matter~\cite{Ohnishi:2011jv}.

Another example of dense asymmetric matter is the neutron star core.
In the neutron star core, the internal temperature is of the order of
$10^6~\mathrm{K}\sim 10^{-4}~\mathrm{MeV}$,
which is small enough compared with the Fermi energy of neutrons.
The baryon density would reach a few times of the nuclear density,
$\sim 10^{15}~\mathrm{g/cm}^3$.
Since the neutron density is much larger than the proton density,
the isospin chemical potential, $\dmu=(\mu_n-\mu_p)/2=(\mu_d-\mu_u)/2$,
is finite and large. In relativistic mean field (RMF) models,
$\dmu$ is calculated to reach 100 MeV in the neutron star core.
Thus we can regard the neutron star core matter as asymmetric matter
at zero temperature.

In Fig.~\ref{fig:PDT0}, we compare the first order phase transition
boundary in PQM 
and $\beta$ equilibrium line
in
RMF at $T=0$~\cite{spion}.
Here we show the boundary for several values of $r$.
RMF parameter sets of
TM1~\cite{Sugahara:1993wz} and IOSTY~\cite{Ishizuka:2008gr}
are adopted as typical examples.
TM1 is a model which describes bulk properties of normal and neutron rich
nuclei as well as the nuclear matter saturation point.
IOSTY is an extended version of TM1, which includes 
degrees of freedom of nucleons and hyperons.
This comparison shows that for $r= 0.2$ and 0.3, large $\dmu$
makes the chiral transition in neutron star crossover.
In IOSTY, 
hyperons are calculated to appear at $\muB \simeq 1100~\mathrm{MeV}$,
then the transition to quark matter occurs before hyperons appear.    
Since the first order transition generally makes the equation of state
softer at around the transition density, the crossover nature
may help to keep the EOS stiff enough
and to support the heavy neutron stars~\cite{Demorest:2010bx}.

We also compare the phase boundary at $T=0$ with those in flavor SU(3) NJL
model results of the neutron star matter~\cite{Abuki:2009ba}.
Since constituent quark mass in Ref.~\cite{Abuki:2009ba} is different
from that of the present work,
we show their results with shifted $\mu_B$. We find that NJL shows
small $\delta\mu$ values around the transition.  This difference mainly
comes from the isovector coupling with quarks and nucleons.
In quark matter, we have chosen the vector coupling in the range $0 \leq r
\leq 0.3$. In nuclear matter, isovector-vector coupling is chosen to
reproduce binding energies of neutron rich nuclei,
and it corresponds to $r \simeq 1.0 \sim 1.2$.
Thus $\delta\mu$ is calculated to be larger in nuclear matter. 
%

\section{Summary}
We have investigated the QCD phase transition
in isospin asymmetric matter using the Polyakov loop extended quark meson (PQM) model.
Specifically, we have discussed 
 isospin chemical potential $\dmu$ and  quark-vector meson coupling dependence of
 the QCD phase boundaries. In PQM,
 we show $\dmu$ reduces the temperature
 of the QCD critical point, and for large $\dmu$, the
 critical point is found to disappear.
 We also show the finite quark-vector meson
 coupling shifts the chiral phase boundary to higher baryon chemical
 potential and reduces the temperature of the CP. This scenario is in agreement
 with the one obtained within other chiral models~\cite{Ohnishi:2011jv,Abuki:2009ba}.

We have also discussed the order of the chiral phase transition in neutron
stars from the comparison of
the QCD phase diagram in PQM and
the $\beta$ equilibrium line in RMF.
In neutron stars, $\dmu$ is large, then the
temperature of the CP becomes lower. Therefore the chiral
phase transition may be crossover, even if
the transition in symmetric matter ($\dmu=0$) is the first order.
In this study, however, we use ($\muB, \dmu$)
values on the $\beta$ equilibrium line
in neutron star matter
calculated with RMF models which do not include the QCD phase
transition effects.
In order to discuss the QCD phase transition in compact
astrophysical phenomena more precisely,
we need the EOS which includes both baryonic and quark degrees of freedom.

One may consider the reduction of $\TCP$ shown in this paper
would contradict
to the finite lepton-number chemical potential result~\cite{PNJL-lepmu},
which suggests the insensitivity of $\TCP$ as a function of the lepton-number
chemical potential. Their results correspond to the $\dmu$ range $\dmu
\lesssim 40~$MeV, while we find that the shift of $T_\mathrm{CP}$ is
large in the range $\dmu \gtrsim 50~$MeV. Thus their results could be
consistent with ours.

The phase diagram structure shown in this article
is based on the assumption that the $s$-wave pion condensation is not 
realized in dense baryonic matter.
This assumption is consistent with the functional renormalization group 
calculation~\cite{FRG} and $s$-wave $\pi N$ repulsion arguments~\cite{spion},
while the results are not in agreement with the mean field results
of PNJL at finite $\dmu$~\cite{PNJL-dmu}.
As a future work, it is an interesting problem to discuss 
the $p$-wave pion condensation, the inhomogeneous chiral condensate,
and the color superconductor phases
in the three-dimensional thermodynamic variable space, $(T,\mu,\dmu)$.

\section*{ACKNOWLEDGMENTS}
T.N. and H.U. are supported by Grants-in-Aid for the Japan Society for
Promotion of Science(JSPS) Research Fellows (Nos. 22-3314 and 25-2148).
This work was supported in part
by Grants-in-Aid for Scientific Research
from the Japan Society for the Promotion of Science (JSPS)
(Nos.
  23340067, 
  24340054, 
  24540271 
  10J03314
),
by Grant-in-Aid for Innovative Areas
from the Ministry of Education, Culture, Sports, Science and Technology
of Japan (MEXT)
(Area No. 2404, Nos. 24105001, 24105008), 
by the Yukawa International Program for Quark-Hadron Sciences,
and by a Grant-in-Aid for the global COE program
``The Next Generation of Physics, Spun from Universality and Emergence''
from MEXT.

\appendix

\section{Critical point within the NJL model at zero temperature\label{A1}}
In the main body of this article we have discussed the effect of an imbalance
of the chemical potentials of $u$ and $d$ quarks on the location of the critical point (CP)
of the QCD phase diagram. Our argument was based mainly on numerical results obtained within
the PQM model. We found that finite $\delta\mu$ moves CP towards a smaller chemical potential
and a lower temperature. Therefore, we might expect that a large enough $\delta\mu$ causes CP
to hit the $T=0$ plane, then disappearing from the phase diagram.
In this Appendix we discuss the same topic within the NJL model. 
We limit ourselves to consider a system of $u$ and $d$ quarks in
the chiral limit: this simplifies the calculations, and allows to identify unambiguously
the location of the chiral phase transition in the phase diagram.
Our purpose is to show analytically how finite $\delta\mu$ induces a softening of the
chiral phase transition at finite $\mu$, pushing the CP to lower values of temperature (and
baryon chemical potential). Eventually, for large enough $\delta\mu$ the CP hits the $T=0$ plane. 
For the purpose of our discussion it is therefore enough to consider the system
at $T=0$ and study the change of the order of the chiral phase transition at finite $\mu$. 

The thermodynamic potential of the NJL model at zero temperature can be written as~\cite{NJL}
\begin{eqnarray}
\Omega &=& \frac{\sigma^2}{G} -2N_c N_f\int\frac{d{\bm p}}{(2\pi)^3}E_p\nonumber\\
&&+2N_c\sum_f\int\frac{d{\bm p}}{(2\pi)^3}(E_p - \mu_f)\Theta\left(\mu_f - E_p\right)~,
\label{eq:A11}
\end{eqnarray}
where $E_p = \sqrt{{\bm p}^2 + M^2}$ with $M=2\sigma = -4 G\langle\bar q_f q_f\rangle$. 
Here $G$ corresponds to the 4-fermion NJL coupling constant, and in agreement with the
notation of the main text we have put $\mu_u = \mu -\delta\mu$ and $\mu_d = \mu + \delta\mu$.
The last addendum on the r.h.s. of the above equation corresponds to the valence quarks contributions.
The vacuum part is regularized by cutting the momentum integral at the scale $|\bm p| = \Lambda$.

Our strategy is as follows: we perform a Ginzburg-Landau expansion of the effective potential,
\begin{equation}
\Omega = \frac{\alpha_2}{2}\sigma^2 + \frac{\alpha_4}{4}\sigma^4 + \frac{\alpha_6}{6}\sigma^6~,
\label{eq:GL_1}
\end{equation} 
where we have subtracted an irrelevant term
which does not depend on the condensate. At zero temperature and finite chemical potential the
coefficients are easily determined from an expansion of Eq.~\eqref{eq:A11} around $\sigma=0$. We get
\begin{eqnarray}
\alpha_2 &=& \frac{2}{G} - \frac{4 N_c}{\pi^2}\Lambda^2 + \frac{2 N_c}{\pi^2}\left(\mu_u^2 + \mu_d^2\right)~,\\
\alpha_4 &=& -\frac{48 N_c}{\pi^2}\left(2-\log\frac{\Lambda^2}{\mu_u \mu_d}\right)~,\label{eq:PPPkk}\\
\alpha_6 &=& \frac{480 N_c}{\pi^2}\left(\frac{1}{\mu_u^2} + \frac{1}{\mu_d^2}\right)~.
\end{eqnarray}
We notice that $\alpha_6 > 0$ causing the potential to be bounded from below. 
As a consequence it is possible to study the phase transition studying the signs
of the first two coefficients. The phase transition is of first (second) order if $\alpha_4 <0$ ($\alpha_4 >0$). 
At the critical point, where the first and second order transition lines meet, one has
$\alpha_2 = \alpha_4 = 0$. Solving $\alpha_2=0$ leads to a relationship between $\mu$ and $\delta\mu$;
using the solution of the latter in the equation $\alpha_4 = 0$ leads to the critical
value of $\delta\mu\equiv\delta\mu_c$ at which the CP hits the $T=0$ plane,
namely
\begin{equation}
\delta\mu_c^2 = -\frac{\pi^2}{2G N_c N_f} + \Lambda^2\left(\frac{1-e^{-2}}{2}\right)~ 
\end{equation}
Using the standard parameters of the model~\cite{NJL} we find $\delta\mu_c \approx 140$ MeV.
This result shows that finite $\delta\mu$ changes the order of the chiral phase transition
at zero temperature and finite chemical potential. 

The fact that finite $\delta\mu$ leads to the softening of the phase transition can be grasped
from Eq.~\eqref{eq:PPPkk}; in fact, for $\delta\mu \ll \mu$ one has
\begin{equation}
\alpha_4 \approx \alpha_4(\delta\mu=0) + \frac{48 N_c}{\pi^2}\frac{\delta\mu^2}{\mu^2}~;
\end{equation} 
the above equation shows that $\delta\mu \neq 0$ makes $\alpha_4$ less negative, thus favoring 
a second order phase transition.The same conclusion can be drawn by
using an extended version of the GL analysis including derivative terms~\cite{GL}.






\end{document}